# Simple setups for quantum games in optical networks

[*]Rubens Viana Ramos   Paulo Benício Melo de Sousa

*Department of Teleinformatic Engineering, Federal University of Ceara, Campus do Pici, 725, C.P. 6007, 60755-640, Fortaleza-Ceará, Brazil*

**Abstract**

In this work, we propose two optical setups for two-players, non-zero and zero sum, quantum games in optical networks using light polarization of single-photon pulses, single-photon detectors and linear optical devices. The optical setups proposed can be easily implemented permitting a fast experimental realization of quantum games with present technology.

Quantum games were introduced by Meyer [1] and Eisert [2] as generalization of classical games [3,4]. Some interesting proposals of implementations of quantum games based on the use of entanglement have been proposed [5-7]. In despite of their correctness, the experimental proposed setups are a little bit complicated to implement. In this work we present two optical setups (one for non-zero and other for zero sum) able to implement quantum games in optical networks using only linear optical devices and devices available with current technology. The non-zero sum game used is the well known two prisoners' dilemma [2-6]. The performance of the setup here proposed for this game is limited by the performance of the single-photon detectors, losses in the optical devices, polarization control in optical networks and efficiency of the single-photon source used. On the other hand, the zero-sum game used is a kind of even-odd game using two entangled states. Since it is not possible to produce entangled states deterministically using only linear optical devices [8-10], the setup for this game has its performance limited by the performance of single-photon source and detectors, losses in optical devices and probability of success of the linear optical setup to produce two entangled states.

___________________
[*]Corresponding author
*E-mail addresses*: rubens@deti.ufc.br, benicio@deti.ufc.br.

The most famous quantum game is the quantum prisoner's dilemma and, since it has been extensively discussed, here we will not be concerned with its details. In an optical implementation of the classical prisoners' dilemma using light polarization, the payoff table could be as shown in Table 1.

Table 1: Payoffs for classical prisoner's dilemma.

| Alice \ Bob | $|H\rangle$ | $|V\rangle$ |
| --- | --- | --- |
| $|H\rangle$ | (2,2) | (5,1) |
| $|V\rangle$ | (1,5) | (4,4) |

Hence, the two strategies are: 1) To use horizontal polarization. 2) To use vertical polarization. For this game, $|H\rangle|H\rangle$ is the Nash equilibrium while $|V\rangle|V\rangle$ is the Pareto equilibrium. On the other hand, the quantum prisoner's dilemma can be played in an optical network using the optical setup shown in Fig. 1.

In Fig. 1 $C_1$ and $C_2$ are optical circulators, M are common mirrors and $L$, $L_1$ and $L_2$ are fiber lengths obeying the relations $L_1 > L$ and $L_2 = (L_1 - L)$. After some calculations, one finds, for the input state $[\alpha|H\rangle + \beta|V\rangle]_\text{I} \otimes [\lambda|H\rangle + \xi|V\rangle]_\text{II}$, the output state

$$|\Psi\rangle_{12} = \alpha\lambda|HH\rangle_{12} + \beta\xi|VV\rangle_{12} + \alpha\xi|HV,0\rangle_{12} + \beta\lambda|0,HV\rangle_{12}. \tag{1}$$

The state $|HH\rangle_{12}$, for example, means a horizontally polarized photon at output 1 and another horizontally polarized photon at output 2, while the state $|0,HV\rangle_{12}$ means none photon at output 1 and two photons, one horizontally and other vertically polarized at output 2. The prisoner's dilemma can be played using the setup of Fig. 1 in the following way: Alice chooses according to her strategy the values of the parameters $\alpha$ and $\beta$ while Bob chooses according to his strategy the values of the parameters $\lambda$ and $\xi$ (equivalently, one can think that Alice chooses the unitary matrix $U_a$ such that $U_a|H\rangle=\alpha|H\rangle+\beta|V\rangle$ and Bob chooses the unitary matrix $U_b$ such that $U_b|H\rangle=\lambda|H\rangle+\xi|V\rangle$). Alice and Bob send their photons (inputs I and II, respectively) and after a known time that depends on the distance between Alice and Bob, they perform a polarization measurement in the optical pulse they have received (outputs 1 and 2, respectively). The possible results of a single measurement in Alice and Bob, according to (1), are: $|HH\rangle$ (having payoff (2,2)), $|VV\rangle$ (having payoff (4,4)), $|0,HV\rangle$ (having payoff (1,5)) and $|HV,0\rangle$ (having payoff (5,1)). The average payoffs are shown in Table 2.

Table 2: Average Payoffs for quantum prisoner's dilemma using setup in Fig. 1.

| Bob<br>Alice | $\lambda|H\rangle+\xi|V\rangle$ | |
|---|---|---|
| $\alpha|H\rangle+\beta|V\rangle$ | $(2|\alpha\lambda|^2,2|\alpha\lambda|^2)$ | $(5|\alpha\xi|^2,1|\alpha\xi|^2)$ |
| | $(1|\beta\lambda|^2,5|\beta\lambda|^2)$ | $(4|\beta\xi|^2,4|\beta\xi|^2)$ |

Considering the table of the average payoffs, the regions where Nash and Pareto equilibriums are equals are the squares (I) $0\leq|\alpha|^2,|\lambda|^2\leq0.443443\ldots$ and (II) $0.600600\leq|\alpha|^2,|\lambda|^2\leq1$. Hence, in those regions, when Alice tries to do the best to herself and Bob tries to do the best for himself, they are doing together the best for the society formed by both.

For the zero-sum game, by its turn, one player wins what the other loses or when a player wins the other necessarily loses, as happens in the even-odd guessing game. A possible quantum implementation of a zero-sum two-player quantum game is presented in Fig. 2 [11], in which the players $A$ and $B$, locally distant, use two pure bipartite entangled states. Initially, player $A$ enters with qubits $U_A|0\rangle \otimes (|0\rangle+|1\rangle)/2^{1/2} = (a|0\rangle+b|1\rangle) \otimes (|0\rangle+|1\rangle)/2^{1/2}$ and player $B$ enters with qubits $U_B|0\rangle \otimes (|0\rangle+|1\rangle)/2^{1/2} = (c|0\rangle+d|1\rangle) \otimes (|0\rangle+|1\rangle)/2^{1/2}$. The choices of $U_A$ and $U_B$ are, respectively, $A$'s and $B$' strategies. The unitary matrix $U_{AB}$ is a four qubit operation providing as output two qubits for player $A$ and two qubits for player $B$. The unitary operator $U_{AB}$ works as follows:

$$U_{AB}\left[a|0\rangle+b|1\rangle\right]_A \otimes \left[\frac{|0\rangle+|1\rangle}{\sqrt{2}}\right]_A \otimes \left[c|0\rangle+d|1\rangle\right]_B \otimes \left[\frac{|0\rangle+|1\rangle}{\sqrt{2}}\right]_B = |\Psi\rangle \quad (2)$$

$$|\Psi\rangle = \left(a|00\rangle+b|11\rangle\right)_{13} \otimes \left(c|00\rangle+d|11\rangle\right)_{24} = \left[ac|0000\rangle+bc|1010\rangle+ad|0101\rangle+bd|1111\rangle\right]_{1234} \quad (3)$$

From (3) one realizes that $A$ and $B$ share two entangled states. The two first qubits (1 and 2) are measured by $A$ using the measurers $M_1^A$ and $M_2^A$, respectively, while and the last two qubits (3 and 4) are measured by $B$ using the measurers $M_1^B$ and $M_2^B$, respectively. Considering these information, a simple game can be implemented using the following rules: player $A$ wins (which implies that $B$ loses) if the results of the measurements in $M_1^A$ and $M_2^A$ are equal (the same happens with the results in $B$'s measurers). Obviously, $B$ wins the game if the values measured by $M_1^B$ and $M_2^B$ are different (the same happen with the results of $A$'s detectors). The probabilities of $A$ and $B$ to win are, hence, given by:

$$P_A = 1 - \left(|a|^2 - 2|a|^2|c|^2 + |c|^2\right) \tag{4}$$
$$P_B = 1 - P_A \tag{5}$$

Now, let us now consider the optical setup shown in Fig. 3 having the input state $(a|H\rangle+b|V\rangle)\otimes(\alpha|H\rangle+\beta|V\rangle)\otimes(c|H\rangle+d|V\rangle)\otimes(\sigma|H\rangle+\delta|V\rangle)$. After some calculations one easily finds the output state is [12]

$$|\Psi\rangle_{1234} = ac\sigma\gamma|HHHH\rangle + ad\sigma\delta|VHVH\rangle + bc\beta\gamma|HVHV\rangle + bd\beta\delta|VVVV\rangle + |\Omega\rangle \tag{6}$$

where $|\Omega\rangle$ is the state containing terms with at least one output with zero photons. Comparing (3) and (6) one sees that, when the quantum operation does not fail (one photon in each output), they are compatible. Moreover, having $c=d=\sigma=\beta=(|0\rangle+|1\rangle)/2^{1/2}$, the output state (6), when it does not fail, is $(\gamma|HH\rangle+\delta|VV\rangle)_{13}\otimes(a|HH\rangle+b|VV\rangle)_{24}$, exactly the type of state required for the realization of the proposed game.

Summarizing, we have proposed two setups for implementation of quantum games in optical networks, one for the quantum prisoners' dilemma and the other for a zero-sum game. Both setups use only linear optical devices, light polarization, single-photon sources and single-photon detectors. Hence, both of them can be experimentally realized with present technology. Their performance is limited by losses in the optical devices (since single-photon pulses are used), efficiency of single-photon source and detectors, random polarization rotations in the optical channel and, in the case of the zero-sum game, the probabilistic behavior of the entangled states generation.

**Acknowledgements**

This work was supported by the Brazilian agency FUNCAP.

**References**

[1] D. Meyer, Phys. Rev. Lett., 82, 5, (1999) 1052.

[2] J.Eisert, M. Wilkens and M. Lewenstein, Phys. Rev. Lett., 83, 15, (1999) 3077.

[3] Iqbal A., Studies in the Theory of Quantum Games, Doctor of Philosophy Thesis. Quaid-i-Azam University, Islamabad, Pakistan (2004).

[4] A. P. Flitney, Aspects of Quantum Game Theory. Doctor of Philosophy Thesis. Department of Electrical and Electronic Engineering, University of Adelaide, Australia. (2005).

[5] J. Du, H. Li, X. Xu, M. Shi, J. Wu, X. Zhou and R. Han, Phys. Rev. Lett., 88, 13, (2002) 137902.

[6] L. Zhou and Le-M. Kuang, Phys. Lett. A, 315, (2003) 426.

[7] J. Du, C. Ju and H. Li, J. Phys. A. Math Gen, 38, (2005) 1559.

[8] E. Knill, R. Laflamme and G. J. Milburn, Nature, 409, (2001) 46.

[9] T. B. Pitman, B. C. Jacobs and J. D. Franson, quant-ph/0404059 (2004).

[10] T. C. Ralph, A. G. White, W. J. Munro, and G. J. Milburn, Phys. Rev. A, 65, (2001) 012314.

[11] P. B. M. de Souza, R. V. Ramos, J. T. C. Filho, quant-ph/0608131 (2006).

[12] J. B. R. Silva, R. V. Ramos, Phys. Lett. A, 359, 6, (2006) 592.

# FIGURE 1

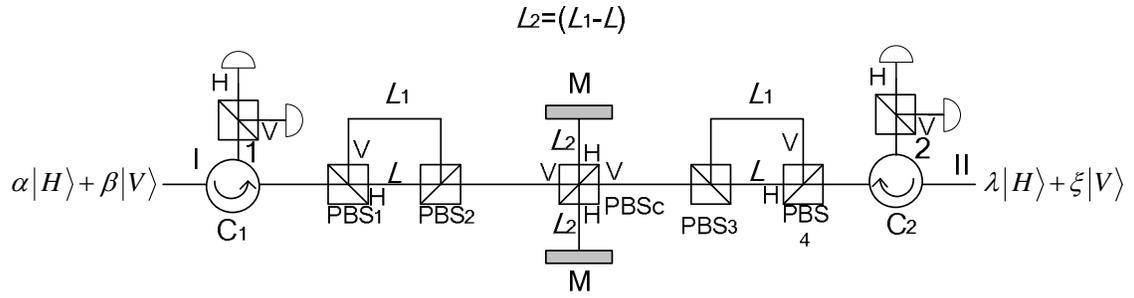

Figure 1: Optical setup for quantum prisoners' dilemma game in an optical network.

# FIGURE 2

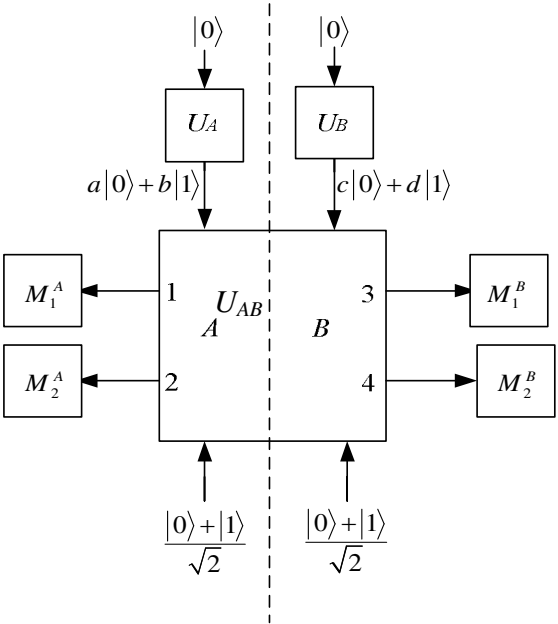

Figure 2: Zero-sum quantum game using two entangled states.

**FIGURE 3**

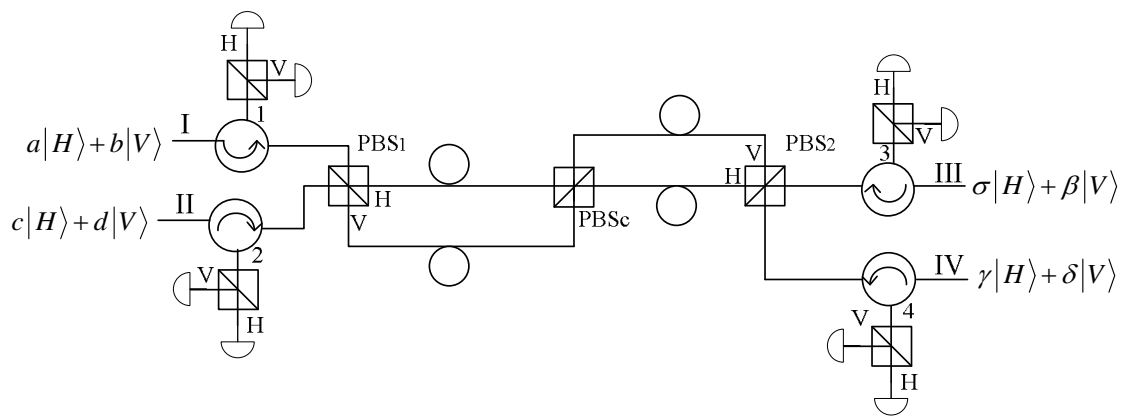

Fig. 3 –Optical setup for zero-sum game.